\documentclass[12pt]{article}
\usepackage{latexsym}
\usepackage{seceqn}
\usepackage{epsf}
\usepackage{othersym}

\newcommand{\mz}{M_Z}
\newcommand{\mt}{m_t}
\newcommand{\mb}{m_b}
\newcommand{\mtau}{m_\tau}
\newcommand{\mgut}{M_\mathrm{GUT}}
\newcommand{\mevf}{M_\mathrm{EVF}}
\newcommand{\msusy}{M_\mathrm{SUSY}}
\newcommand{\mcomp}{M_\mathrm{comp}}
\newcommand{\agut}{\alpha_\mathrm{GUT}}
\newcommand{\ygut}{Y_\mathrm{GUT}}
\newcommand{\yt}{Y_t}
\newcommand{\yb}{Y_b}
\newcommand{\ytau}{Y_\tau}
\newcommand{\yu}{Y_U}
\newcommand{\yd}{Y_D}
\newcommand{\ye}{Y_E}
\newcommand{\yubar}{Y_{\bar{U}}}
\newcommand{\ydbar}{Y_{\bar{D}}}
\newcommand{\yebar}{Y_{\bar{E}}}
\newcommand{\R}{R_{b\tau}}
\newcommand{\GeV}{\mathrm{GeV}}
\newcommand{\TeV}{\mathrm{TeV}}
\newcommand{\MSbar}{\overline{\mathrm{MS}}}
\newcommand{\tgut}{t_\mathrm{GUT}}
\newcommand{\tsusy}{t_\mathrm{SUSY}}
\newcommand{\nvector}{n_\mathrm{vector}}
\newcommand{\andvol}[3]{{\bf #1}~(#3)~#2}
\newcommand{\PRL}[3]{Phys.~Rev.~Lett.~\andvol{#1}{#2}{#3}}
\newcommand{\PRD}[3]{Phys.~Rev.~\andvol{D#1}{#2}{#3}}

\newcommand{\NPB}[3]{Nucl.~Phys.~\andvol{B#1}{#2}{#3}}
\newcommand{\PLB}[3]{Phys.~Lett.~\andvol{B#1}{#2}{#3}}

\newcommand{\ZPC}[3]{Z.~Phys.~\andvol{C#1}{#2}{#3}}

\newcommand{\PTP}[3]{Prog.~Thoer.~Phys.~\andvol{#1}{#2}{#3}}

\newcommand{\MPLA}[3]{Mod.~Phys.~Lett.~\andvol{A#1}{#2}{#3}}

\newcommand{\hepph}[1]{\ [hep-ph/#1]}
\newcommand{\hepth}[1]{\ [hep-th/#1]}
\newcommand{\KEK}[1]{\ [KEK Scanned Preprint #1]}
\begin{document}

\begin{titlepage}

\renewcommand{\thefootnote}{\fnsymbol{footnote}}

\begin{flushright}
CERN--TH/96--363 \\
ICCR-Report-379-97-2 \\
hep-ph/9612493
\end{flushright}

\medskip

\begin{center}

\textbf{\Large
Predictions of $\mathbf{\mb/\mtau}$ and $\mathbf{\mt}$\\
in an Asymptotically Non--Free Theory}

\bigskip
\smallskip

\textsc{Masako BANDO}
\\
\textit{Aichi University, Miyoshi, Aichi 470-02, Japan}\footnote{%
E-mail: bando@aichi-u.ac.jp}

\medskip

\textsc{Tetsuya ONOGI}
\\
\textit{Department of Physics,
Hiroshima University, Hiroshima 606-01, Japan}\footnote{%
E-mail: onogi@theo.phys.sci.hiroshima-u.ac.jp}

\medskip

\textsc{Joe SATO}
\\
\textit{Institute for Cosmic Ray Research \\ 
The University of Tokyo,
Midori-Cho, Tanashi, Tokyo 188, Japan}\footnote{%
E-Mail: joe@icrhp3.icrr.u-tokyo.ac.jp}

\medskip

\textsc{Tatsu TAKEUCHI}
\\
\textit{CERN, TH Division, CH--1211 Gen\`eve 23, Switzerland}\footnote{%
E-mail: takeuchi@vxcern.cern.ch}

\bigskip

\begin{abstract}
We discuss an extension of the
Minimal Supersymmetric Standard Model (MSSM)
with 5 generations of matter superfields.
The extra generations are assumed to form a
generation--mirror generation pair (the 4th and anti-4th generations)
enabling the extra fermions to have $SU(2)_L\times U(1)_Y$ invariant masses.
Due to the contribution of the extra generations,
all three running gauge couplings of 
$SU(3)_C \times SU(2)_L \times U(1)_Y$ become asymptotically non--free
while preserving gauge coupling unification at the GUT scale.  
We show that due to the asymptotically non--free character of the
gauge couplings: 
(1) the top and bottom Yukawa couplings are strongly focused onto 
infrared fixed points as they are evolved down in scale
making their values at $\mu=\mz$ insensitive 
to their initial values at $\mu=\mgut$;
(2) the model predicts 
$\R(\mz) \equiv Y_b/Y_\tau |_{\mu =\mz}\approx 1.8$, 
which is consistent with the experimental value
provided we take the ratio of Yukawa couplings at the GUT scale to be
$\R(\mgut) = Y_b/Y_\tau |_{\mu = \mgut} = 1/3$;
(3) the $t$ mass prediction comes out to be $\mt\approx 180\,\GeV$
which is also consistent with experiment.
\end{abstract}

\end{center}

\vfill


\end{titlepage}

\renewcommand{\thefootnote}{\arabic{footnote}}
\setcounter{footnote}{0}


\section{Introduction}

The popularity of the Minimal Supersymmetric Standard Model (MSSM) in recent 
years is mainly due to its success in attaining gauge coupling unification:
given the particle content of the MSSM, the three coupling constants of the 
$SU(3)_C \times SU(2)_L \times U(1)_Y$ gauge groups converge to a common value
at a common scale (the GUT scale) when evolved up to higher energies using 
the renormalization group equations (RGE) \cite{AMALDI:91}.
This unification of the gauge coupling constants is crucial if one wishes
to construct a Grand Unified Theory (GUT) which unifies the three gauge
groups of the Standard Model (SM) into a larger simple group at a single scale.
However, it should be noted that the particle content which achieves
such unification is not unique \cite{HALL:91}.
In particular, as pointed out in Ref.~\cite{IBANEZ:81},
one always has the freedom to add complete generations
of matter superfields to the MSSM 
without destroying the unification condition.\footnote{Of course, if one
adds too many generations, the gauge couplings will reach the Landau pole
before reaching the GUT scale.  See Ref.~\cite{MOROI:93}.}
 
Another attractive feature of the MSSM is the possibility of 
unifying the $b$ and $\tau$ Yukawa couplings:
if one assumes
\begin{equation}
\R(\mgut) = \yb(\mgut)/\ytau(\mgut) = 1
\label{UNI1}
\end{equation}
at the GUT scale\footnote{Whether the condition $\yb(\mgut) = \ytau(\mgut)$
is realized or not in GUT's depends on the representation of the Higgs field
which gives mass to the fermions.
For $SU(5)$, $SO(10)$, and $E_6$ unifications, the Higgs must be  
in the $\mathbf{5}$, $\mathbf{10}$, and $\mathbf{27}$ representations,
respectively.},
then one finds that the MSSM can reproduce the experimental value of 
$\R(\mz) = \yb(\mz)/\ytau(\mz) \approx 1.8$
if the Yukawa couplings of the top and the bottom were such that
$\yt(\mgut) \agt 2 \gg \yb(\mgut)$ ($1 \alt \tan\beta \alt 3$), or
$\yt(\mgut) \approx \yb(\mgut) \approx 1$ 
($40 \alt \tan\beta \alt 60$).\footnote{
Note that since $\mt/\mb = (\yt/\yb)\tan\beta$,
the region $\yt \gg \yb$ corresponds to
small $\tan\beta$ while $\yt \approx \yb$ corresponds to large $\tan\beta$.
The lower and upper limits of $1 \alt \tan\beta$ and
$\tan\beta \alt 60$ are required to keep $\yt$ and $\yb$ in the perturbative
region throughout evolution between $\mz$ and $\mgut$.}
\cite{LANGACKER:94}

The reason why the experimental value of $\R$ can only be 
reproduced for either small or large $\tan\beta$ is easy 
to understand\footnote{We assume the reader has 
some familiarity with the RGE's for the Yukawa couplings}:
QCD interactions will enhance $\yb(\mu)$ over $\ytau(\mu)$
as they are evolved down from $\mgut$ to $\mz$ so that
$\R(\mz)$ will end up well above the experimental value if
only running due to gauge interactions were taken into account.
This QCD effect must be countered by strong Yukawa interactions which
will slow down the running.   
A smaller value of $\R$ consistent
with experiment can be obtained when $\yt$ is large enough to
counter the QCD enhancement alone, or when both $\yt$ and $\yb$
are large so that the two of them combined can have the desired effect.
In the intermediate $\tan\beta$ region ($3 \alt \tan\beta \alt 40$)
$\yt$ is not large enough to sufficiently suppress the 
increase of $\R$ by itself while $\yb$ is not large enough to
compensate for it.

Of these two solutions, the small $\tan\beta$ case is 
often considered particularly attractive since 
the large size of
$\yt(\mgut)$ will drive $\yt(\mu)$ rapidly towards an
infrared \textit{quasi}--fixed point \cite{HILL:81} 
as it is evolved down in scale. 
As a result, the value of $\yt(\mz)$ is highly insensitive to its
initial value $\yt(\mgut)$ at the GUT scale.
On the other hand, the large $\tan\beta$ case opens the
possibility of unifying the top Yukawa coupling with the
other two:
\begin{equation}
\yt(\mgut) = \yb(\mgut) = \ytau(\mgut),
\label{UNI2}
\end{equation}
as required in $SO(10)$ unification with a $\mathbf{10}$--Higgs.
However, the insensitivity to the initial condition at $\mgut$ is
lost.\footnote{Another problem with the large $\tan\beta$ solution is
that fine tuning of the Higgs potential is necessary to achieve radiative
electroweak symmetry breaking.    In the small $\tan\beta$ case,
radiative electroweak symmetry breaking is naturally achieved due to the
initial condition $\yt(\mgut) \gg \yb(\mgut)$.  However, fine tuning
is necessary in this case also to obtain the correct value of $\tan\beta$.
See, for instance, Ref.~\cite{BANDO:93}.}

In this paper, we wish to outline how these conclusions will be 
modified when the MSSM is extended with an addition of 
a generation--mirror generation pair of extra matter superfields. 
(the 4th and $\bar{4}$th generations)\footnote{
Such pairs are well known to exist in many GUT scenarios.
See Ref.~\cite{GEORGI:79}.}.
Each generation is assumed to
consist of the usual 15 chiral fermion fields plus their superpartners.  
We will ignore the right--handed neutrino
necessary to form the $\mathbf{16}$ representation of $SO(10)$ since
we will always assume it to have a superheavy Majorana mass and make it 
decouple from the RGE's.\footnote{We do not consider an intermediate scale
for the right--handed neutrino mass for the sake of simplicity.
See Refs.~\cite{VISSANI:94,BRIGNOLE:94} for analyses of the MSSM case with an
intermediate scale.}
Due to the mirror quantum number assignments between the   
4th and $\bar{4}$th generation fermions, they can develop
$SU(2)_L\times U(1)_Y$ invariant masses enabling the left--handed neutrino
to have a heavy Dirac mass thus circumventing the LEP limit for the
number of massless neutrinos.
Also, radiative corrections to LEP
observables from the extra fermions can be made to decouple by making
this gauge invariant mass large.\cite{MAEKAWA:95}

One immediate consequence of the presence of the 2 extra generations
is that all three gauge couplings of $SU(3)_C\times SU(2)_L\times U(1)_Y$ 
will be asymptotically non--free:
they will become larger as they are evolved up to 
coincide at the unification scale~\cite{THEISEN:88}.
This property is actually unique to the 5 generation model.
In models with 4 generations\footnote{%
4 generation models have been discussed in Refs.~\cite{FOURFAMILY}.}
or less, the QCD coupling will stay
asymptotically free, and in models with 6 generations or more
the couplings will diverge before unification.

As shown in the appendix, the unification of gauge couplings
is controlled solely by the \textit{differences} of 
the beta function coefficients in the one-loop approximation. 
Since the differences of the coefficients 
are independent of the number of full generations, the gauge
coupling unification in our 5 generation model works well just as in
the MSSM.

\begin{figure}[tp]
\centering
\unitlength=1cm
\begin{picture}(10,8)
\unitlength=1mm
\put(1,77){$\alpha$}
\put(50,41){$\beta > 0$}
\put(60,15){$\beta < 0$}
\put(12,2){$\mz$}
\put(77,2){$\mgut$}
\put(7,35){Experimentally}
\put(7,30){allowed range}
\put(82,12){Fine Tuning}
\put(82,52){No Fine Tuning}
\ \epsfbox[90 160 349 377]{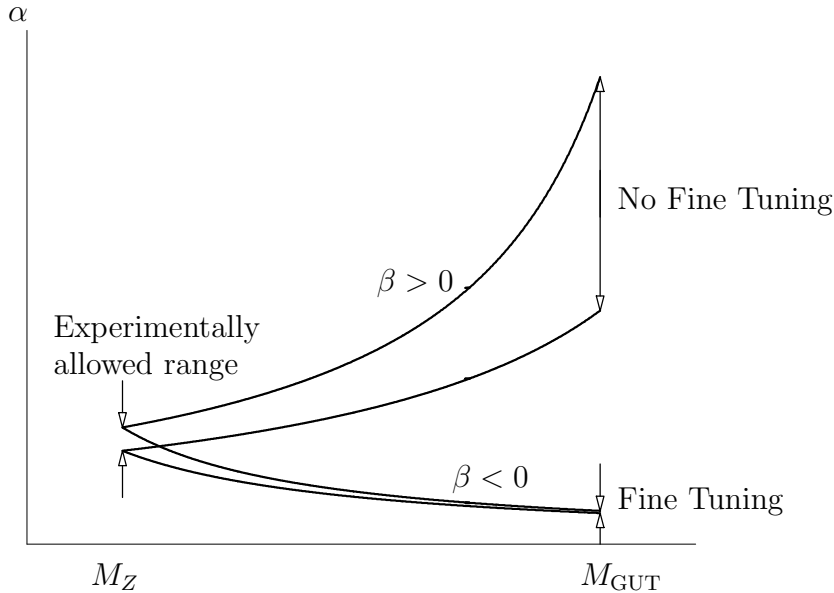}
\end{picture}
\vspace{-5pt}
\caption[]{%
The difference between asymptotically free and non--free theories.
}
\label{fig0}
\end{figure}

However, an important difference between asymptotically free theories and
asymptotically non--free theories is that $\alpha = 0$ is an
ultraviolet (UV) fixed point in the former but
an infrared (IR) fixed point in the latter. 
This means that for asymptotically free (non--free) theories,
the RG flow will be such that a large region of $\alpha$ values
in the IR (UV) will flow into a small region close 
to $\alpha=0$ in the UV (IR),
and the difference in the relative size of these regions will be more
pronounced for larger separations in scale.
Therefore, in order to get the desired value of $\alpha(\mz)$
in asymptotically free theories, the value of $\alpha(\mgut)$
must be tuned to extreme accuracy while for  
asymptotically non--free theories, no fine tuning is necessary.
This is shown schematically in Fig.~\ref{fig0}.

This absence of the necessity to fine tune parameters at the
UV cutoff is an extremely attractive feature of asymptotically non--free
theories.   It means that the high energy theory effective above the
UV cutoff can give the correct predictions at low energies as long as it
predicts the values of the running couplings at the cutoff to be
within an only mildly restricted range.
This point has been emphasized previously by many 
authors \cite{PARISI:75} (though not necessarily from a modern point of view).
In particular, Moroi, Murayama and Yanagida \cite{MOROI:93}
have studied the same 5 generation model as we are considering here
and have shown that the values of the running couplings at $\mgut$
need not even be unified to predict the correct value of
$\sin^2\theta_w$.



In this paper, we extend the analysis of Moroi et al. and study 
how the existence of the extra generations will affect the
running of the Yukawa couplings of the 3rd generation fermions.
A similar problem for the non--supersymetric case has been considered
in \cite{GRUN:88}.
As in the MSSM case, we will impose a unification condition on the
Yukawa couplings at $\mgut$ and determine the parameter range in which 
our model can predict the correct top, bottom and $\tau$--lepton masses.

The attentive reader at this point may think that such a program is doomed
to failure from the beginning.
Since the QCD coupling is asymptotically non--free, the QCD enhancement
of $\R$ from $\mgut$ to $\mz$ will be even larger than the MSSM case
making it impossible to bring $\R(\mz)$ down to $\sim 1.8$
even with large Yukawa couplings.
However, we would like to quickly point out that 
the unification condition need not be that of 
Eqs.~(\ref{UNI1}) or (\ref{UNI2}).
In fact, an $SO(10)$--GUT with an $\mathbf{126}$--Higgs predicts
\cite{CHANOWITZ:77}
\begin{equation}
\yt(\mgut) = \yb(\mgut) = \frac{1}{3}\ytau(\mgut),
\label{UNI3}
\end{equation}
so that $\R(\mgut) = 1/3$.  
This is the unification condition which we will adopt.\footnote{%
An $\mathbf{126}$--Higgs is necessary to to give a direct 
Majorana mass term to the right--handed neutrino.
}
In this case, the extra enhancement from
QCD is actually welcome since $\R$ must be enhanced by a factor of 
$5 \sim 6$ to reproduce the experimental value of $\R(\mz)$.

This paper is organized as follows:
In section~2, we describe our model and specify the way we calculate
the RG evolution of the gauge and Yukawa couplings.
In section~3, we show how the gauge couplings can be unified in our
model.   Section~4 discusses Yukawa coupling unification and the
predictions for $\R(\mz)$ and $\mt$.
Section~5 concludes.


\section{The $4+\bar{1}$ Generation Model:} 

In extending the MSSM by introducing extra matter superfields, we
must keep two things in mind:
(1) the matter superfields must be introduced in such a way that
gauge coupling unification (and anomaly cancellation)
of the MSSM is preserved, and
(2) the fermion content must be compatible with the constraints
placed by LEP measurements, namely three massless neutrino species
and the so--called Peskin--Takeuchi constraint \cite{PESKIN:90}.

The simplest way to satisfy these requirements is to
introduce 2 extra generations which form a generation--mirror
generation pair.  
We will call them the 4th and anti--4th generations.
The fermion content of these extra generations will be
`vector--like' so that all of them, including the extra neutrinos, 
can develop $SU(2)_L\times U(1)_Y$ invariant Dirac masses.
These masses will also suppress the size of radiative 
corrections to LEP observables from the extra fermions enabling
them to circumvent the Peskin--Takeuchi constraint\cite{MAEKAWA:95}.

It should be noted that we can only introduce one such
generation--mirror generation pair.
If we introduce two pairs or more, all three gauge couplings will 
reach their Landau poles and diverge well before the would--have--been
unification scale\cite{MOROI:93}.

We denote the extra fermion families $(U,D,N,E)$ and 
$(\bar{U},\bar{D},\bar{N},\bar{E})$, respectively, and
give them a common $SU(2)_L \times U(1)_Y$ invariant
mass of $\mevf$.
Their superpartners, and all the other supersymmetric particles
in the theory will be given a common SUSY breaking mass of $\msusy$.
For the sake of simplicity, we take $\mevf = \msusy = 1\,\TeV$.

In addition to the $SU(2)_L \times U(1)_Y$ invariant masses,
we also couple the 4th and $\bar{4}$th generation fermions to the
two Higgs doublets in the same way as the other generations.
Here we take the case where 
\begin{eqnarray}
\yu = \yt,  \qquad \yubar = 0, \cr
\yd = \yb,  \qquad \ydbar = 0, \cr
\ye = \ytau,\qquad \yebar = 0.
\label{yukawas}
\end{eqnarray}
and set all the 1st and 2nd generation Yukawa couplings to zero.
Furthermore, we impose the unification condition
\begin{equation}
\yt(\mgut) = \yb(\mgut) = \frac{1}{3} \ytau(\mgut) \equiv \ygut,  
\end{equation}
as mentioned in the introduction.

In view of the relatively large coupling strengths near the 
unification scale due to the asymptotically non-freeness, 
we use the fully coupled 
2--loop renormalization group equations (RGE's)
from Ref.~\cite{RGE:2LOOP} to evolve the gauge and Yukawa 
couplings.
We ignore small differences in the masses of the 
4th and $\bar{4}$th generation
particles or that of the supersymmetric particles 
which may be induced by the Yukawa couplings and simply set all
their masses at $\mevf = \msusy = 1\,\TeV$.
We also ignore threshold corrections.
Therefore, between $\mgut$ and $\mevf = \msusy$, 
we evolve the couplings with
the RGE's for the Supersymmetric SM with 5 super--generations
and 2 super--Higgs doublets, while 
between $\mevf = \msusy$ and $\mz$, we use the RGE's for the SM
with only 3 ordinary generations and 1 Higgs doublet.
The gauge couplings are connected continuously at 
$\mevf = \msusy = 1\,\TeV$ while the up--type (down--type) Yukawa 
couplings are multiplied by $\sin\beta$ ($\cos\beta$) below $\mevf = \msusy$
to take into account the decoupling of one of the Higgses.

The number of adjustable parameters in our model is four:
the unification scale $\mgut$, 
the unified gauge coupling $\agut$,
the unified Yukawa coupling $\ygut$, 
and the mixing angle of the low lying Higgs fields
$\tan\beta = v_2/v_1$, where $\sqrt{v_1^2 + v_2^2} = v = 246\,\GeV$.
We restrict $\agut$ and $\ygut$ to the region
\begin{equation}
\agut < 1.0, \qquad \ygut < 0.7.
\label{eq:rangealy}
\end{equation}
(Note that $\ytau(\mgut) = 3\ygut$.
Note also that the natural expansion coefficient corresponding to
the $\alpha_i(\mu)$'s is $Y^2/(4\pi)$ for the Yukawa's.)
As we will see later, this will keep
the gauge and Yukawa couplings within their
perturbative regions throughout the evolution from $\mgut$ to $\mz$.

Since we do not consider the evolution of the soft SUSY breaking
parameters of the Higgs potential in this paper,
$\tan\beta$ will remain a phenomenological parameter to be fixed by hand.
We will use the $\tau$--lepton mass to fix $\tan\beta$ from
\begin{equation}
\mtau(\msusy) = \frac{v}{\sqrt{2}}\ytau(\msusy)\cos\beta.  
\end{equation}



\section{Gauge Coupling Unification:}

\begin{figure}[tp]
\centering
\unitlength=1cm
\begin{picture}(14,9)
\unitlength=1mm
\put(4,78){$\agut$}
\put(124,5){$\log_{10}\mgut(\GeV)$}
\epsfxsize=12cm
\ \epsfbox{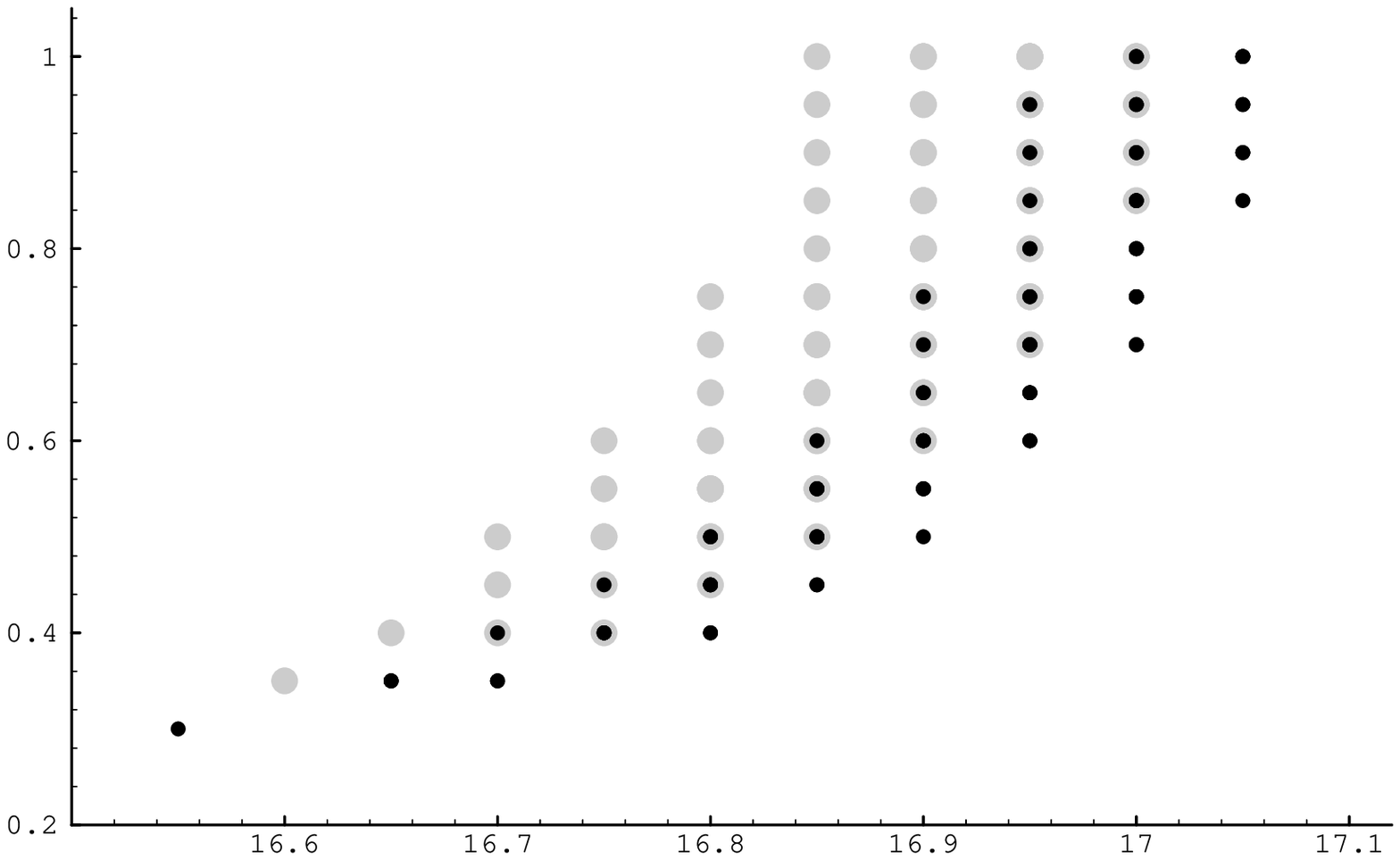}
\end{picture}
\vspace{-5pt}
\caption[]{
The allowed region in the plane $(\agut,\mgut)$. 
The small--black, and large--gray circles indicate the ranges 
$0.4 \le \ygut < 0.7$, and 
$0.1 < \ygut < 0.4$, respectively.
}
\label{fig:Z1}
\end{figure}

\begin{figure}[tp]
\centering
\unitlength=1cm
\begin{picture}(14,9)
\unitlength=1mm
\put(25,76){$\alpha_1^{-1}$}
\put(25,45){$\alpha_2^{-1}$}
\put(25,22){$\alpha_3^{-1}$}
\put(124,5){$\log_{10}\mu(\GeV)$}
\epsfxsize=12cm
\ \epsfbox{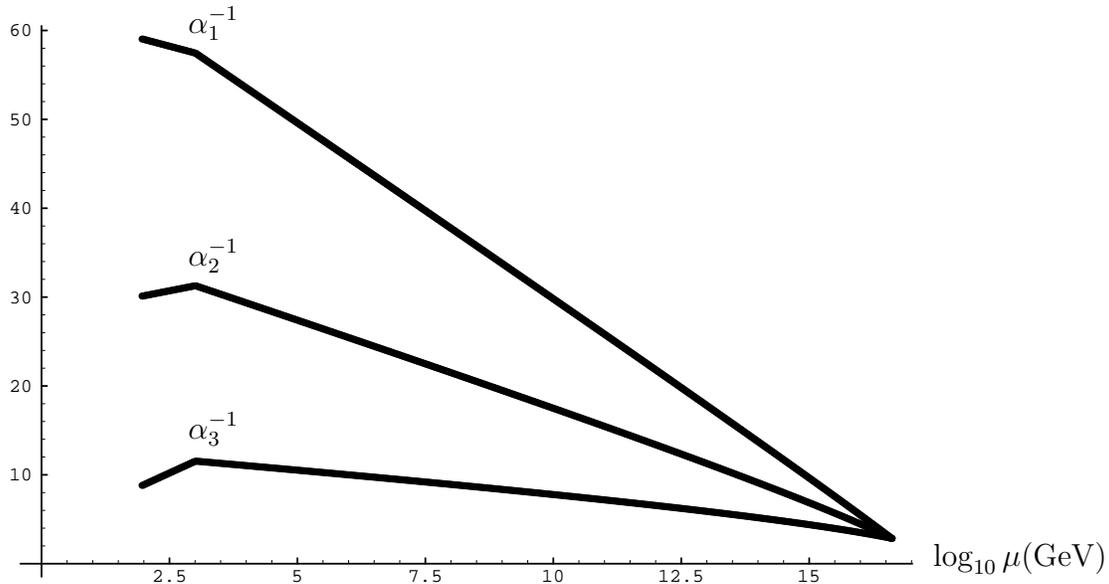}
\end{picture}
\vspace{-5pt}
\caption[]{
Typical $\mu$ dependence of 
$\alpha_1(\mu),\alpha_2(\mu)$ and $\alpha_3(\mu)$. 
The parameter values for this plot were 
$(\agut,\ygut,\mgut,\tan\beta) = (0.35,\,0.3,\,10^{16.6}\GeV,\,57.5)$.}
\label{fig:Z2}
\end{figure}

The values of the $SU(3)_C\times SU(2)_L\times U(1)_Y$ coupling constants
at $\mu = \mz$ are given by 
\cite{PDG:96} :
\begin{eqnarray}
\alpha_1(\mz) & = & 0.01689           \pm 0.00005,\\
\alpha_2(\mz) & = & 0.03322           \pm 0.00025, \\
\alpha_3(\mz) & = & 0.12\phantom{000} \pm 0.01.
\label{eq:expalpha}
\end{eqnarray}
Note that these are the $\MSbar$ coupling constants\footnote{%
This may a confusing point since the `effective' QED coupling constant
$\alpha(\mz)$ and the `effective' weak angle $\sin^2\theta_W$ that
are usually quoted by LEP are not $\MSbar$ values.}
and that the $U(1)$ coupling is normalized to
$\alpha_1 = (5/3)g^{\prime 2}/4\pi$.

For fixed values of $\ygut$ in the range given in Eq.~(\ref{eq:rangealy}),
we searched for values of $\agut$ and $\mgut$ which reproduced 
the experimental data given above.
The results are shown in Fig.~\ref{fig:Z1}.

We see that the allowed range of $\agut$ is narrow 
for smaller $\mgut$ but still exist down to $\mgut \approx 10^{16.55}\GeV$
and becomes wider as $\mgut$ is increased.
This result is as expected from our discussion on asymptotically non--free
theories:  a wider range of $\agut$ corresponds to a much smaller range
of couplings at $\mz$, and the allowed range will become wider as $\mgut$
is increased.
However, if we increase $\mgut$ beyond $\sim 10^{17.1}\GeV$, then
$\agut$ and/or $\ygut$ will have to be taken beyond the limits specified
in Eq.~(\ref{eq:rangealy}) and they will be too large for the perturbative
treatment of the RGE's to be reliable.

As an example, we show the running of the three gauge couplings 
in Fig.~(\ref{fig:Z2}) for typical values of $\agut$, $\mgut$, and $\ygut$.
We see a small deviation from linear dependence on $\log\mu$ near
$\mgut$ where the couplings become large and the two--loop
corrections start contributing to the running appreciably.
However it is clear that two--loop contributions are still not very
serious within the range of $\agut$ which we have chosen here and 
we may regard our perturbative treatment to be sufficient.


\section{Yukawa Coupling Unification:}

\begin{figure}[tp]
\centering
\unitlength=1cm
\begin{picture}(14,8.5)
\unitlength=1mm
\put(4,78){$\ytau$}
\put(124,5){$\log_{10}\mu(\GeV)$}
\epsfxsize=12cm
\ \epsfbox{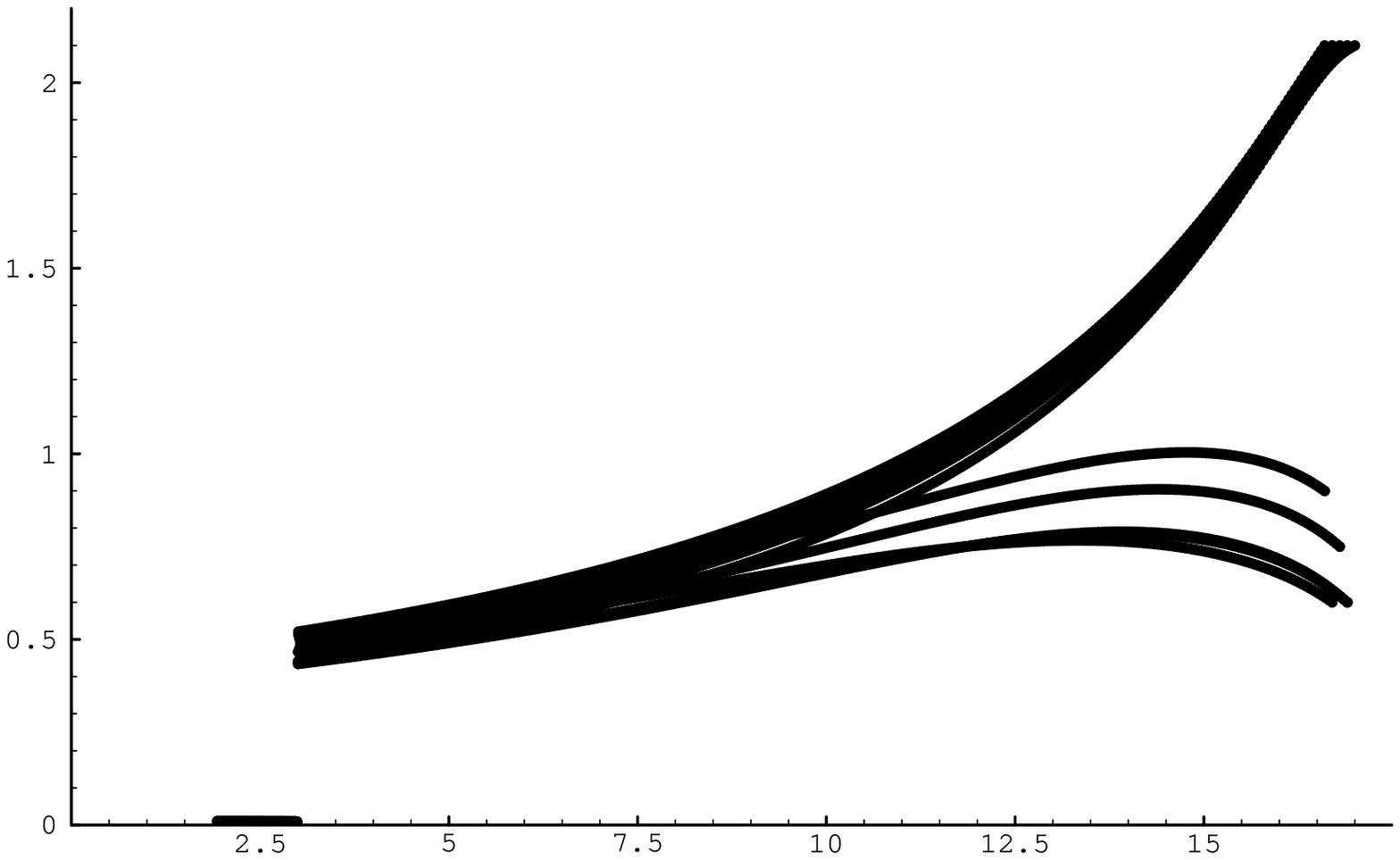}
\end{picture}
\vspace{-5pt}
\caption[]{%
The running of the $\tau$--lepton Yukawa coupling for
typical values of $(\agut, \mgut)$ shown in Fig.~\ref{fig:Z1}.
The value of $\agut$ is varied between 0.3 and 0.8 while
$\ytau(\mgut) = 3\ygut$ is varied between 0.6 and 2.1.
The Yukawa coupling is multiplied by $\cos\beta$ below $\msusy$.
}
\label{fig:ytau}
\end{figure}

\begin{figure}[tp]
\centering
\unitlength=1cm
\begin{picture}(14,8.5)
\unitlength=1mm
\put(4,78){$\yb$}
\put(124,5){$\log_{10}\mu(\GeV)$}
\epsfxsize=12cm
\ \epsfbox{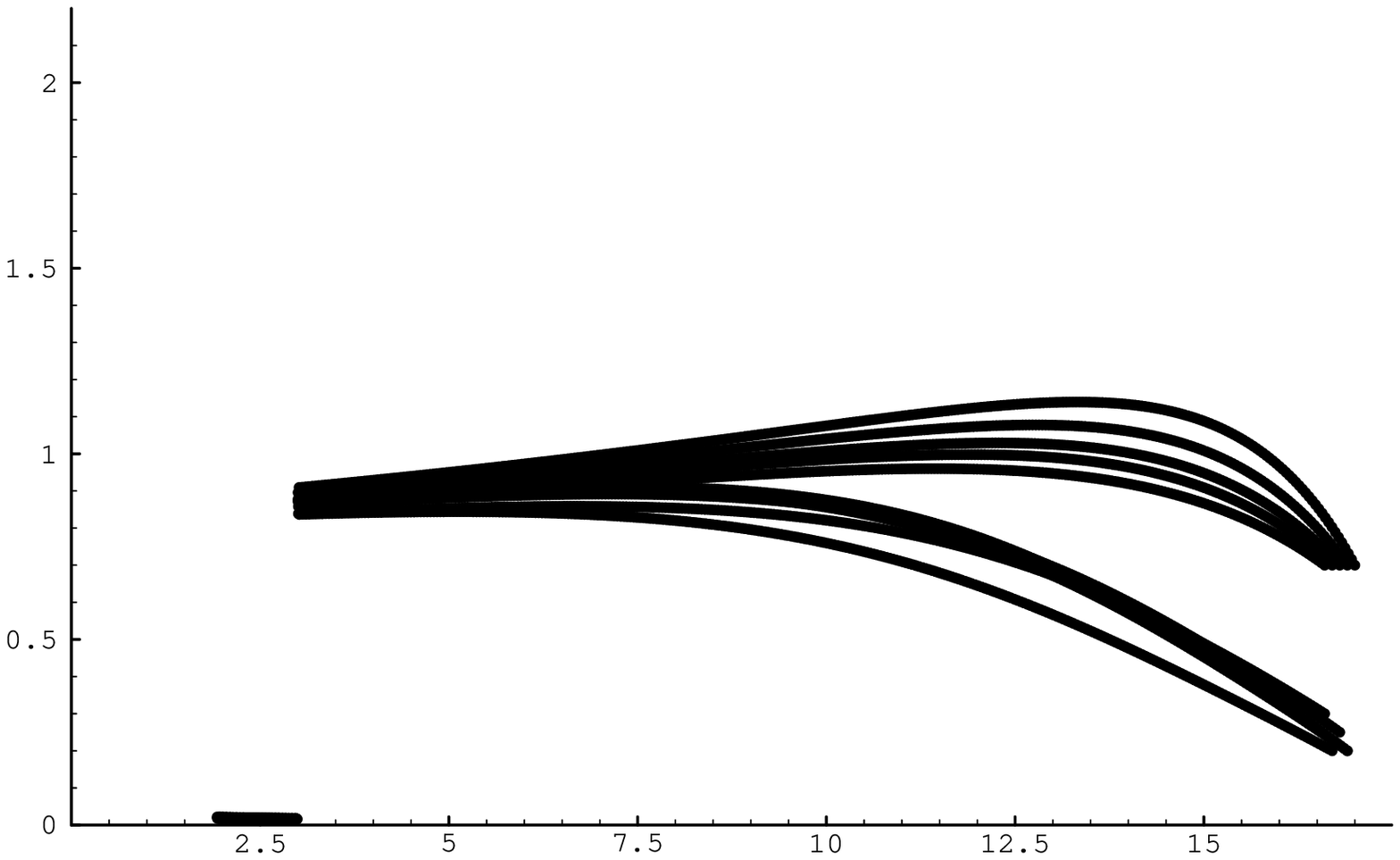}
\end{picture}
\vspace{-5pt}
\caption[]{%
The running of the bottom Yukawa coupling for
typical values of $(\agut, \mgut)$ shown in Fig.~\ref{fig:Z1}.
The value of $\agut$ is varied between 0.3 and 0.8 while
$\yb(\mgut) = \ygut$ is varied between 0.2 and 0.7.
The Yukawa coupling is multiplied by $\cos\beta$ below $\msusy$.
}
\label{fig:ybottom}
\end{figure}

\begin{figure}[tp]
\centering
\unitlength=1cm
\begin{picture}(14,8.5)
\unitlength=1mm
\put(4,76){$\yt$}
\put(124,2){$\log_{10}\mu(\GeV)$}
\epsfxsize=12cm
\ \epsfbox{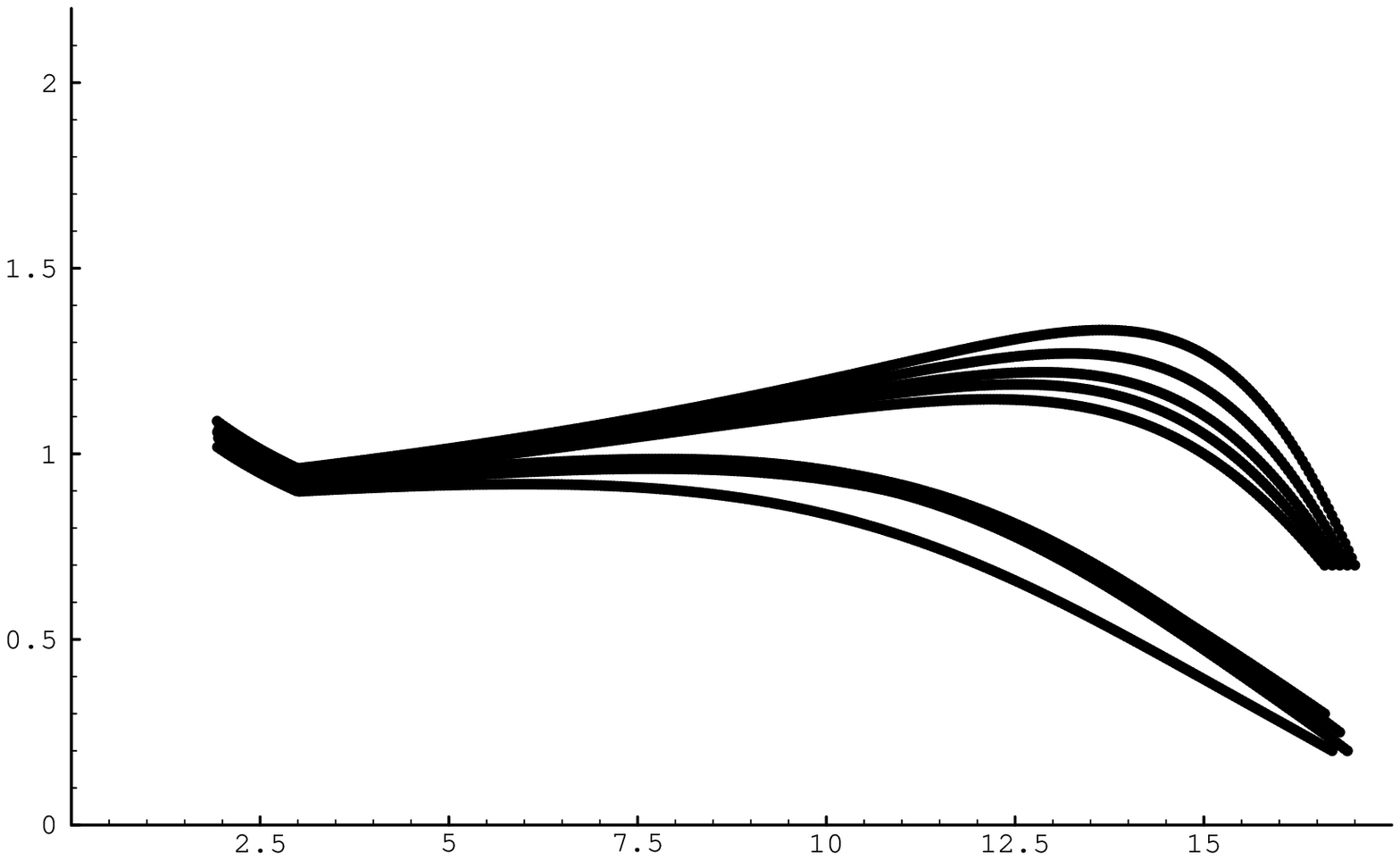}
\end{picture}
\vspace{-5pt}
\caption[]{%
The running of the top Yukawa coupling for
typical values of $(\agut, \mgut)$ shown in Fig.~\ref{fig:Z1}.
The value of $\agut$ is varied between 0.3 and 0.8 while
$\yt(\mgut) = \ygut$ is varied between 0.2 and 0.7.
The Yukawa coupling is multiplied by $\sin\beta$ below $\msusy$.
}
\label{fig:ytop}
\end{figure}

Next, we fix the values of $\agut$ and $\mgut$ in the
range allowed by gauge coupling unification and calculate the evolution of
the Yukawa couplings for different values of $\ygut$.
Typical evolutions of the $\tau$, $b$, and $t$ Yukawa couplings are
shown in Figs.~\ref{fig:ytau}, \ref{fig:ybottom}, and \ref{fig:ytop}.
As is evident from these figures, the asymptotic non--freeness of the
gauge couplings has a strong focusing effect on the top and bottom 
Yukawa couplings
as they evolve down in scale and as a result, 
the values of the two Yukawa's converge to IR fixed points
by the time they reach the SUSY breaking scale $\msusy = 1\,\TeV$.
In the case of the $\tau$ Yukawa coupling, the situation is rather
different. Near the GUT scale it tends to focus itself
due to its larger size at $\mgut$ (Recall $\ytau(\mgut) = 3\ygut$.).
However, unlike the top and bottom Yukawa couplings, 
once $\ytau$ becomes small at lower scales, it runs 
slowly and does not quite converge to its IR fixed point ($y_{\tau}=0$).

Due to this IR fixed point behavior, the values of $\yt(\mz)$, 
$\yb(\mz)$ have almost no dependence on $\ygut$. They do depend on
the value of $\agut$ but even then only very mildly.


\subsection{Bottom to Tau Mass Ratio:}

\begin{figure}[tp]
\centering
\unitlength=1cm
\begin{picture}(14,9)
\unitlength=1mm
\put(4,78){$\R(\mz)$}
\put(110,30){$\log_{10}\mgut(\GeV)$}
\epsfxsize=12cm
\ \epsfbox{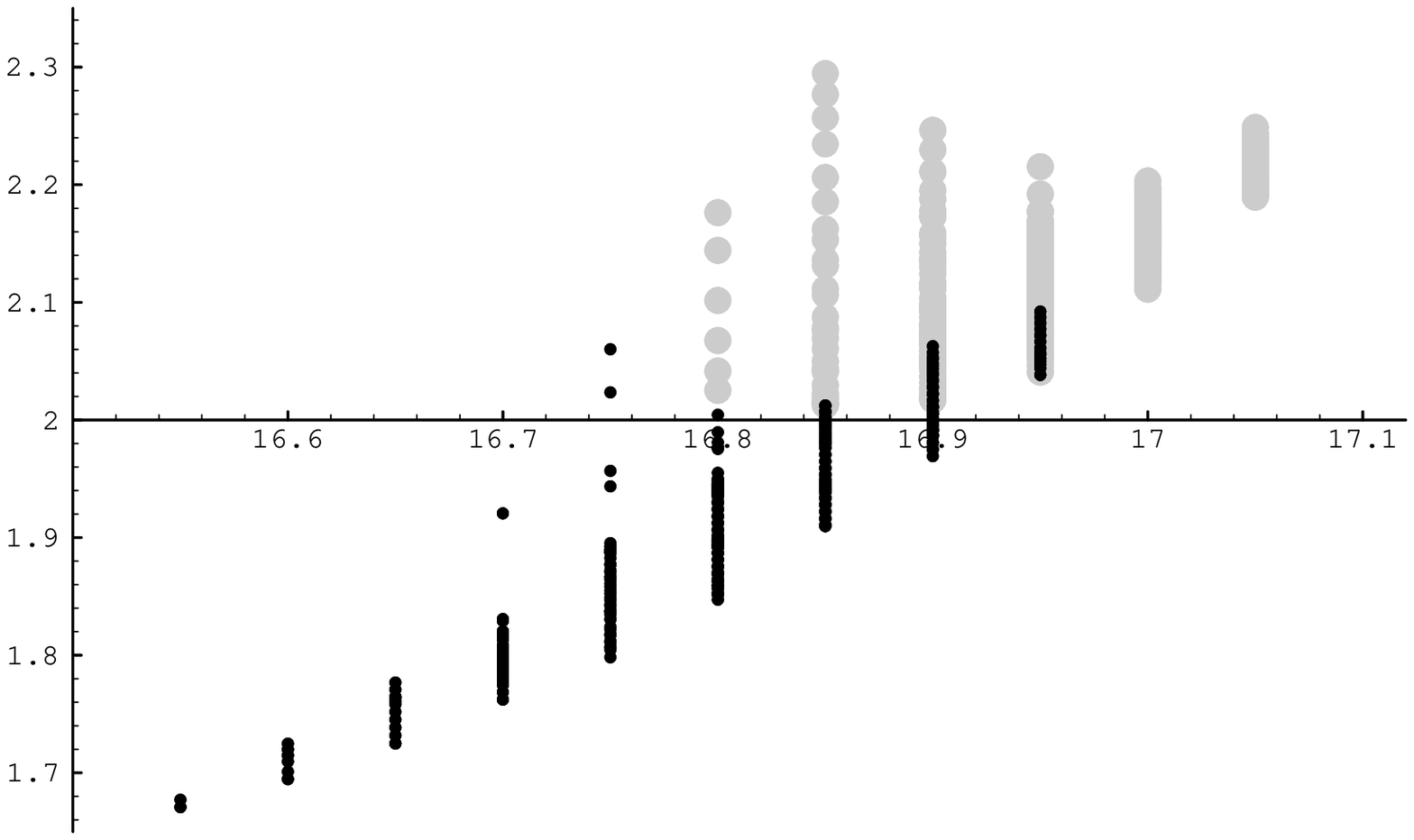}
\end{picture}
\vspace{-5pt}
\caption[]{
The dependence of $\R(\mz)$ on $\mgut$ and $\agut$. 
The small--black, and large--gray circles indicate the ranges 
$0.3 < \agut \le 0.6$, and 
$0.6 < \agut < 1$, respectively.
The dependence of $\R(\mz)$ on $\ygut$ is negligible.
}
\label{fig:Rbtau}
\end{figure}

The $b$--$\tau$ mass ratio has been the most intensively studied
quantity in both supersymmetric and non--supersymmetric GUT scenarios.
Many authors have investigated the possibility of unifying the two
couplings with various degrees of success.
\cite{HALL:91,LANGACKER:94,VISSANI:94,
BRIGNOLE:94,KELLEY:92,BARGER:93,BARDEEN:94,BANDO:94,POKORSKI:95}

Currently, the experimentally determined $\MSbar$ running masses 
of the $\tau$--lepton and the $b$ quark at $\mu = \mz$ are given by
\cite{PDG:96}
\begin{eqnarray}
\mtau(\mz) & = & 1.75 \pm 0.01 \,\GeV, \cr
\mb(\mz)   & = & 3.1  \pm 0.4  \,\GeV,
\end{eqnarray}
%
from which we conclude
\begin{equation}
\R(\mz) = 1.6 \sim 2.0
\label{expRbtau}
\end{equation}



The dependence of $\R(\mz)$ on $\agut$ and $\mgut$ in our model is
shown in Fig.~\ref{fig:Rbtau}.
Obviously, 
whether our model can reproduce the experimental value of $\R(\mz)$
or not depends almost solely on the value of $\agut$.
If $\agut > 0.6$, then 
the QCD interactions near $\mgut$ will be so strong that $\yb$ will
be enhanced too much relative to $\ytau$.
However, there is still a large set of $(\agut, \mgut)$ values
which keeps $\R(\mz)$ below 2.

Of course, since $\tan\beta$ is large in our model
there is potentially a very large radiative correction to $\mb$
from the loop induced coupling of the $b$ quark to $v_2$.  
\cite{HALL:91,BARDEEN:94,BANKS:88,HEMPFLING:94}
This correction can throw our prediction off the mark by
a considerable amount.

However, such corrections can easily be compensated for.
If the correction makes $\R(\mz)$ smaller, we only need to make $\agut$
larger.   If it makes $\R(\mz)$ larger, we only need to make $\agut$
smaller, changing $\mevf$ and/or $\msusy$ if necessary.

We therefore conclude that our model can accommodate the $b$--$\tau$ 
mass ratio quite easily without any fine tuning.


\subsection{The Top Quark Mass:}

Due to the IR fixed point behavior of 
$\ytau$, $\yb$, and $\yt$ we have seen 
above, for the range of $(\alpha, \mgut, \ygut)$ values that
yield the correct value of $\R(\mz)$ we find:  
\begin{equation}
\yt(\mz) = 1.01 \sim 1.07.
\end{equation}
Using the $\tau$ mass to fix $\tan\beta$, we find
\begin{equation}
\mt(\mz) = 176 \sim 187 \,\GeV,
\end{equation}
which is in perfect agreement with the  
experimental value 
\cite{PDG:96} :
\begin{equation}
\mt(\mz) = 180 \pm 10 \GeV.
\end{equation}
This result is actually highly dependent on our choice
Eq.~(\ref{yukawas}) for the 4th and $\bar{4}$th generation Yukawa
couplings.
Had we chosen a different condition such as
\begin{eqnarray}
\yu = \yubar = \yt, \cr
\yd = \ydbar = \yb, \cr
\ye = \yebar = \ytau,
\label{yukawa2}
\end{eqnarray}
then the IR fixed value for $\yt$ would have been
\begin{equation}
\yt(\mz) \sim 0.705
\end{equation}
leading to a prediction of
\begin{equation}
\mt(\mz) \sim 154 \GeV
\end{equation}
%


\section{Summary and Conclusions :}

We have presented an extension of the MSSM with
a generation anti--generation pair of extra matter superfields.
The $SU(3)_C\times SU(2)_L\times U(1)_Y$ gauge couplings are
asymptotically non--free in this model but still
converge to a common value before any of them diverge.
Consequences of this asymptotically non--free theory are:
\begin{enumerate}

\item
The unified coupling $\agut$ and the unification scale $\mgut$
need not be fine tuned to reproduce the values of 
the $SU(3)_C \times SU(2)_L \times U(1)_Y$ gauge couplings at
$\mu=\mz$.
Larger values of $\mgut$ allow for a larger range of $\agut$ values.
However, $\mgut$ must be taken below $\sim 10^{17.1}\GeV$ to keep
$\agut$ and $\ygut$ within perturbative range.

\item 
The top and bottom Yukawa couplings are strongly 
focused onto IR fixed points.
This makes the IR values of the two Yukawa couplings insensitive to
their initial value $\ygut$ at the GUT scale.

\item
The model can reproduce the ratio of bottom and tau
masses $\R(\mz) = \yb(\mz)/\ytau(\mz) =  1.6 \sim 2$, 
provided we assume the unification condition
$\R(\mgut) = 1/3$ ($\mathbf{126}$--Higgs) instead of the usual 
$\R(\mgut) = 1$ ($\mathbf{10}$--Higgs).

\item
The model can also give the correct top mass 
provided the Yukawa couplings of the 4th and $\bar{4}$th generations
are taken as in Eq.~(\ref{yukawas}).

\end{enumerate}
It would be most interesting if this insensitivity of the low energy
predictions to the initial conditions at the GUT scale could be taken
further to include the soft SUSY breaking parameters of the Higgs sector.
In particular, if a large $\tan\beta$ could be generated without
fine tuning, it could provide an answer to the question why the
top is so much heavier than the bottom.

Above $\mgut$, our model is a supersymmetric $SO(10)$ theory
which includes four $\mathbf{16}$--plets and an 
$\mathbf{\overline{16}}$--plet, and a Higgs sector which consist of at least
an $\mathbf{126}$ and an $\mathbf{\overline{126}}$ (in order for 
$\mathbf{126}$ to have a mass term).
This makes the $SO(10)$ gauge coupling asymptotically non--free also.

It has recently been argued that
such asymptotic non--freeness of the gauge couplings can be consistently
interpreted as a sign of compositeness. \cite{BANDO:96}
The basic reasoning is as follows:
The general compositeness condition of gauge bosons
is given by $Z(\mcomp)=0$ where $Z(\mu)$ is the wave--function
renormalization constant and $\mcomp$ is the compositeness scale.
If one enforces conventional normalization $Z(\mu) = 1$ at 
all scales $\mu$, then superficially the running gauge coupling
$\alpha(\mu)$ will diverge at $\mu = \mcomp$
making it look like an asymptotically non--free theory.
This is analogous to theories with dynamically generated Higgs bosons
where compositeness manifests itself as the divergence of the 
Yukawa coupling at the compositeness scale in the effective 
Higgs--Yukawa theory ~\cite{BARDEEN:90} .

In the SM, the large number of arbitrary parameters
have lead most people to believe that going beyond the SM will somehow 
reduce the number of parameters and make theories more predictive.  
However, most extensions of the SM such as the MSSM or Technicolor 
actually \textit{increases} the number of parameters by a huge amount.    
Reduction of the number of parameters is usually achieved by imposing 
ad hoc symmetries such as $R$--parity,
universality of the scalar masses at the GUT scale, etc.    
What our analysis has shown is that for certain types of theories
with IR fixed points, it may happen that
most of the parameters simply \textit{do not matter}
or only needs to be specified to an order of magnitude to make precise
low energy predictions.   
Clearly, the IR fixed point phenomena is an alternative to
symmetries for making theories more predictive and deserves
thorough and systematic investigation.


\section*{Acknowledgments}

We would like to thank T.~Kugo and T.~Yanagida for their valuable comments 
and encouragement.
Helpful discussions with W.~A.~Bardeen, M.~Carena, K.-I.~Izawa,
T.~Takahashi, Y.~Taniguchi, S.~Tanimura, and C.~E.~M.~Wagner
are also gratefully acknowledged.
The early stage of this work began when one of the authors (M.B.)
was staying at Fermilab as a summer visitor in 1995 to  
collaborate with T.O. and T.T..  
We would like to thank the members of the Fermilab Theory Group 
for providing us with this opportunity 
and for their warm hospitality during our stay.
The main part of this work was completed 
during the Kashik\=ojima Workshop held in August 1996 at
the Kashik\=ojima Center, Ag\=o-ch\=o, Shima-gun, Mi\'e-ken, Japan,
which was supported by the Grant--in Aid for Scientific Research 
(No.~07304029) from the Japanese Society for the Promotion of Science
(JSPS).
M.B. is supported in part by the Grant--in Aid for
Scientific Research (No.~33901-06640416) also from the JSPS.

\newpage

\appendix


\section{Solution to the one--loop RG equations}

In this appendix, we give a qualitative interpretation
of our results in the one-loop approximation.

The one-loop renormalization group equations above the
SUSY scale in our model are as follows: 
\begin{eqnarray}
\frac{d\alpha_i}{dt} & = & \frac{b_i}{2\pi} \alpha_i^2
\qquad ( i=1,2,3) \label{eqn:alpha}\\
\frac{d\alpha_t}{dt} & = & \frac{\alpha_t}{2\pi} 
\left[ - \left( \frac{13}{15}\alpha_1 + 3 \alpha_2 +\frac{16}{3} \alpha_3 
         \right)
       + ( 9 \alpha_t + \alpha_b) 
\right]   \label{eqn:yukawat}\\
\frac{d\alpha_b}{dt} & = & \frac{\alpha_b}{2\pi} 
\left[ - \left( \frac{7}{15}\alpha_1 + 3 \alpha_2 +\frac{16}{3} \alpha_3 
         \right)
       + ( \alpha_t + 9 \alpha_b + 2 \alpha_{\tau}) 
\right]   \label{eqn:yukawab}\\
\frac{d\alpha_{\tau}}{dt} & = & \frac{\alpha_{\tau}}{2\pi} 
\left[ - \left( \frac{9}{5}\alpha_1 + 3 \alpha_2 
         \right)
       + ( 6 \alpha_b + 5 \alpha_{\tau}) 
\right].  \label{eqn:yukawatau}
\end{eqnarray}
Here, $t=\log\mu$,  $\alpha_{t,b,\tau} \equiv \frac{Y_{t,b,\tau}^2}{4\pi}$, 
and the one-loop beta function coefficients are given by the following 
formula:
\begin{eqnarray*}
(b_1,b_2,b_3) 
& = & (0,-6,-9)+(3+2\nvector)(2,2,2)
       +2\left( \frac{3}{10},\frac{1}{2},0 \right)  \cr
&   & \qquad\qquad\qquad\,\,\;
      \mbox{(for the MSSM plus $\nvector$ full generation pairs)} \cr
& = & \left( \frac{53}{5},5,1
      \right) \qquad
      \mbox{(for $\nvector = 1$)}. 
\end{eqnarray*}

Since the model is nothing but the standard model below the SUSY scale,
as can be seen from Fig.~\ref{fig:Z2}, the experimental inputs are essentially
equivalent to 
\begin{displaymath}
\alpha_1^{-1}(\tsusy) \approx 58,\qquad
\alpha_2^{-1}(\tsusy) \approx 31,\qquad
\alpha_3^{-1}(\tsusy) \approx 12.
\end{displaymath}
The solution to Eq.(\ref{eqn:alpha}) is
\begin{displaymath}
\frac{1}{\alpha_i(t)} 
= \frac{b_i}{2\pi}(\tgut-t)+\frac{1}{\alpha_i(\tgut)} 
\qquad (i=1,2,3).
\end{displaymath}
From the above equation, and using $\alpha_1^{-1}(\tsusy)$ and 
$\alpha_2^{-1}(\tsusy)$
as inputs, we can predict $ \alpha_3^{-1}(\tsusy)$ as
\begin{displaymath}
\alpha_3^{-1}(\tsusy)
= \frac{b_3-b_2}{b_1-b_2}
\left[ \alpha_1^{-1}(\tsusy)- \alpha_2^{-1}(\tsusy) 
\right]
+ \alpha_2^{-1}(\tsusy) \approx 12.
\end{displaymath}
Note that only the differences of beta function coefficients appear
in this expression. Therefore, at the one-loop level, the prediction of 
$\alpha_3^{-1}(\tsusy)$ would be exactly the same for any $\nvector$.
However, the two-loop correction which is $O(\max(\alpha^2(t)))$ would
be different depending on whether the theory is asymptotically free or
non-free. In the MSSM, the expected correction would be
as large as $O(\alpha^2(\tsusy)) \sim 1\%$, whereas in the
$\nvector = 1$ case, the correction would be as large as 
$O(\alpha^2(\tgut)) \sim 10\%$.  Thus if the one-loop prediction
differs from the experiment by more than a few percent, one has to
consider rather large threshold correction at the GUT scale to explain
the discrepancy in the MSSM whereas in the latter model there is
still room for two-loop corrections to account for it.

Next, let us solve Eqs.~(\ref{eqn:yukawat})$\sim$(\ref{eqn:yukawatau}) to 
obtain the low energy behavior of the Yukawa couplings.
The first term in each equation is the contribution from the gauge
interactions and the second term in each equation is that from the Yukawa
interactions. The gauge interactions try to enhance the Yukawa
couplings as the scale is lowered whereas the Yukawa interactions tend
to reduce it.  Therefore, one can naively expect that the Yukawa couplings 
fall into infrared fixed points where the gauge contribution and the
Yukawa contribution balance.  Whether this is true or not depends on 
initial values and the details of the beta function coefficients.

In order to see this more explicitly, 
let us make the following three assumptions and reduce 
Eqs.~(\ref{eqn:yukawat})$\sim$(\ref{eqn:yukawatau}) into a much simpler form:
\begin{enumerate}
\item Since at low energy, $\alpha_1,\alpha_2, \ll \alpha_3$,     
      we can neglect $\alpha_1,\alpha_2$.
\item At low energy, the contribution from $\alpha_{\tau}$ is not so
      dominant compared to those from $\alpha_{t,b}$. 
      This is partly because the coefficient of $\alpha_{\tau}$ in 
      Eq.~(\ref{eqn:yukawab}) is not so large and partly because in
      Eq.~(\ref{eqn:yukawatau}) there is no contribution from
      $\alpha_3$ which can prevent $\alpha_{\rm yukawa}$ getting small 
      thus $\alpha_{\tau}$ gets small at lower energy much faster than
      $\alpha_{t,b}$.   
\item Assuming 1 and 2, the difference between $\alpha_{t}$ and $
      \alpha_{b}$ is almost negligible. This is because we impose 
      $\alpha_t(\tgut) = \alpha_b(\tgut)$
      as the GUT scale initial condition, and because the approximate 
      RG equation is symmetric under the interchange 
      $\alpha_t \leftrightarrow \alpha_b$.
\end{enumerate}
In the following,
we will obtain the solutions to the resulting approximate equations.
Of course, the behavior of those
solutions will be correct only qualitatively since
the corrections from the neglected terms are not completely negligible. 
(In principle, it is possible to check the validity of this
approximation by solving the full equations.) 
However, since we are only
interested in the qualitative features, we will not discuss the 
corrections from the neglected terms here.

With these assumptions and by setting $\alpha_t=\alpha_b\equiv\alpha_Q$,
Eqs.~(\ref{eqn:yukawat}), (\ref{eqn:yukawab}) become
\begin{equation}
\frac{d\alpha_Q}{dt}  =  \frac{\alpha_Q}{2\pi} 
\left( - \frac{16}{3} \alpha_3   +  10 \alpha_Q 
\right).
\label{eqn:yukawaQ}
\end{equation}
Let us define the ratio $z_Q= \frac{\alpha_Q}{\alpha_3}$.
Eq.~(\ref{eqn:yukawaQ}) is then
\begin{equation}
\frac{dz_Q}{dt} = \frac{z_Q}{2\pi} \alpha_3 
\left( - \frac{19}{3}  +  10 z_Q 
\right).
\label{eqn:zQ}
\end{equation}
It is easy to see that the solution to Eq.~(\ref{eqn:zQ}) is
\begin{equation}
\frac{z_Q(t)-\frac{19}{30}}{z_Q(t)} =
\frac{z_Q(\tgut)-\frac{19}{30}}{z_Q(\tgut)}
\exp\left[ -\frac{19}{3}
           \log\left(\frac{\alpha_3(\tgut)}{\alpha_3(t)}
               \right)
    \right]
\label{eqn:zQsol}
\end{equation}

The exponent 
$\frac{19}{3}\log\left(\frac{\alpha_3(\tgut)}{\alpha_3(t)}\right)$
at $t = \tsusy$ ranges roughly from 8 to 12, thus at $\tsusy$ 
the quark Yukawa coupling gets very close to the infrared fixed point.
\begin{equation}
\alpha_Q(t)
\stackrel{t \rightarrow \tsusy}{\Longrightarrow}
\frac{19}{30}\alpha_3(\tsusy)
\label{eqn:alphaQlimit}
\end{equation}
This behavior of the quark Yukawa coupling is consistent with what we 
see from the numerical solution to the two--loop RG equation.

On the other hand, the one--loop RG equation for the bottom--tau ratio
$\R$ can be obtained from
Eqs~.(\ref{eqn:yukawab}), (\ref{eqn:yukawatau}).
Using the assumptions, the equation at low energy can be simplified to
\begin{equation}
\frac{d R_{b\tau}}{dt} 
= \frac{R_{b\tau}}{2\pi} 
\left( -\frac{8}{3} \alpha_3 + 2 \alpha_Q 
\right)
\approx \frac{R_{b\tau}}{2\pi} 
\left( -\frac{7}{5} \alpha_3  
\right).
\label{eqn:Requation}
\end{equation}
It is easy to see that the solution to Eq.~(\ref{eqn:Requation}) is
\begin{equation}
\R(t) = 
\R(\tgut) 
\exp\left[ \frac{7}{5} \log\left(\frac{\alpha_3(\tgut)}{\alpha_3(t)}
                           \right)
    \right]
\label{eqn:Rsol}
\end{equation}
The factor
$\exp\left[ \frac{7}{5}\log\left(\frac{\alpha_3(\tgut)}{\alpha_3(t)}
\right) \right]$
at $t = \tsusy$ is about 4.1 to 7.5 for $\alpha_{\rm GUT}=0.3 \sim 0.55$.
This gives roughly the right enhancement for $\R$.

On the other hand,
the RG equations in the MSSM are given by:
\begin{eqnarray}
\frac{d\alpha_i}{dt} & = & \frac{b_i}{2\pi} \alpha_i^2
\qquad\qquad ( i=1,2,3) \label{eqn:alphas}\\
\frac{d\alpha_t}{dt} & = & \frac{\alpha_t}{2\pi} 
\left[ - \left( \frac{13}{15}\alpha_1 + 3 \alpha_2 +\frac{16}{3} \alpha_3 
         \right)
       + ( 6 \alpha_t + \alpha_b ) 
\right]   \label{eqn:yukawats}\\
\frac{d\alpha_b}{dt} & = & \frac{\alpha_b}{2\pi} 
\left[ - \left( \frac{7}{15}\alpha_1 + 3 \alpha_2 +\frac{16}{3} \alpha_3 
         \right)
       + ( \alpha_t + 6 \alpha_b + \alpha_{\tau} )
\right]   \label{eqn:yukawabs}\\
\frac{d\alpha_{\tau}}{dt} & = & \frac{\alpha_{\tau}}{2\pi} 
\left[ - \left( \frac{9}{5}\alpha_1 + 3 \alpha_2 
         \right)
       + ( 3 \alpha_b + 4 \alpha_{\tau} ) 
\right].  \label{eqn:yukawataus}
\end{eqnarray}
The one--loop beta function coefficients are given by the following 
formula:
\begin{displaymath}
(b_1,b_2,b_3)  =  \left( \frac{33}{5},1,-3 \right).
\end{displaymath}
The equation for $z_Q$ is
\begin{equation}
\frac{dz_Q}{dt} = \frac{z_Q}{2\pi} \alpha_3 
\left( - \frac{19}{3}  +  7 z_Q 
\right).
\label{eqn:zQs}
\end{equation}
It is easy to see that the solution to Eq.~(\ref{eqn:zQs}) is
\begin{equation}
\frac{z_Q(t)-\frac{19}{21}}{z_Q(t)}
=
\frac{z_Q(\tgut)-\frac{19}{21}}{z_Q(\tgut)}
\exp\left[ -\frac{19}{9}
            \log\left( \frac{\alpha_3(t)}{\alpha_3(\tgut)}
                \right)
    \right]
\label{eqn:zQsols}
\end{equation}
The exponent 
$\frac{19}{9}\log\left(\frac{\alpha_3(t)}{\alpha_3(\tgut)}\right)$
at $t = \tsusy$ is roughly 2, thus at $\tsusy$ 
the focusing effect of quark Yukawa coupling is not as strong as in
our 5 generation model.

\newpage

\newcommand{\brk}{\hfill\\}


\begin{thebibliography}{99}

\bibitem{AMALDI:91}
%
%
U.~Amaldi, W.~de~Boer and H.~F\"urstenau, \PLB{260}{447}{1991}, \brk
P.~Langacker and N.~Polonsky, \PRD{52}{3081}{1995} \hepph{9503214}.


\bibitem{HALL:91}
%
%
A.~Giveon, L.~J.~Hall, and U.~Sarid, \brk
\PLB{271}{138}{1991}, \KEK{9109088} \brk
%
%
L.~J.~Hall, R.~Rattazzi, and U.~Sarid, \brk
\PRD{50}{7048}{1994} \hepph{9306309}, \brk
R.~Rattazzi and U.~Sarid, \brk
\PRD{53}{1553}{1996} \hepph{9505428}, \brk
U.~Sarid, UND--HEP--96--US02 \hepph{9610341}.


\bibitem{IBANEZ:81}
L.~Ib\'a\~nez and G.~Ross, \PLB{105}{439}{1981}, \brk
W.~J.~Marciano and G.~Senjanovic, \PRD{25}{3092}{1981}.


\bibitem{MOROI:93}
T~.Moroi, H.~Murayama and T.~Yanagida, \brk
\PRD{48}{R2995}{1993} \hepph{93062268}.

\bibitem{LANGACKER:94}
P.~Langacker and N.~Polonsky, \brk 
\PRD{49}{1454}{1994} \hepph{9306205}; \brk
\PRD{50}{2199}{1994} \hepph{9403306}, \brk
N.~Polonsky, \PRD{54}{4537}{1996} \hepph{9602206}.


\bibitem{HILL:81}
%
%
B.~Pendleton and G.~G.~Ross, \PLB{98}{291}{1981}, \brk
C.~T.~Hill, \PRD{24}{691}{1981}.

\bibitem{BANDO:93}
%
%
M.~Bando, T.~Kugo, N.~Maekawa, and H.~Nakano, \brk
\MPLA{7}{3379}{1992} \KEK{9212293}. \brk
M.~Bando, N.~Maekawa, H.~Nakano, and J.~Sato, \brk
\MPLA{8}{2729}{1993} \KEK{9405421}.


\bibitem{GEORGI:79}
H.~Georgi, \NPB{156}{126}{1979}.

\bibitem{VISSANI:94}
%
%
F.~Vissani and A.~Y.~Smirnov, \PLB{341}{173}{1994} \hepph{9405399}.

\bibitem{BRIGNOLE:94}
%
%
A.~Brignole, H.~Murayama and R.~Rattazzi, \brk
\PLB{335}{345}{1994} \hepph{9406397}.

\bibitem{MAEKAWA:95}
%
%
L.~Lavoura and J.~P.~Silva, \brk
\PRD{47}{2046}{1993}, \KEK{9302379} \brk
N.~Maekawa, \PTP{93}{919}{1995} \hepph{9406375}; \brk
\PRD{52}{1684}{1995} \KEK{9504209}; \brk
KUNS--1366 \hepph{9510414}.


\bibitem{THEISEN:88}
S. Theisen, N.D. Tracas and G. Zoupanos, \ZPC{37}{597}{1988}

\bibitem{FOURFAMILY}
J.~Bagger, S.~Dimopoulos, and E.~Masso, 
\NPB{253}{397}{1985}, \brk
J.~E.~Bj\"orkman and D.~R.~T.~Jones, 
\NPB{259}{533}{1985}, \brk
M.~Cveti\v{c} and C.~R.~Preitschopf, 
\NPB{272}{490}{1986}, \brk
J.~F.~Gunion, D.~W.~McKay, and H.~Pois, \brk
\PLB{334}{339}{1994} \hepph{9406249}, \brk
\PRD{53}{1616}{1996} \hepph{9507323}, \brk
M.~Carena, H.~E.~Haber, and C.~E.~M.~Wagner, \brk
\NPB{472}{55}{1996} \hepph{9512446}.


\bibitem{PARISI:75}
%
%
G.~Parisi, \PRD{11}{909}{1975}, \brk
L.~Maiami, G.~Parisi and R.~Petronzio, \NPB{136}{115}{1978}, \brk
N.~Cabbibo and C.R.~Farrar, \PLB{110}{107}{1982}.

\bibitem{GRUN:88}
G~.Grunberg, \brk
\PRD{38}{R1012}{1988}; \brk
\PLB{203}{R413}{1988}.


\bibitem{CHANOWITZ:77}
%
%
M.~S.~Chanowitz, J.~Ellis and M.~K.~Gaillard, \brk
\NPB{128}{506}{1977}.

\bibitem{PESKIN:90}
%
%
M.~E.~Peskin and T.~Takeuchi, \brk
\PRL{65}{964}{1990}, \PRD{46}{381}{1992}, \brk
J.~L.~Hewett, T.~Takeuchi, and S.~Thomas, \brk
SLAC--PUB--7088, CERN--TH/96--56 \hepph{9603391}.



\bibitem{RGE:2LOOP}
V.~Barger, M.~S.~Berger, and P.~Ohmann, \brk
\PRD{47}{1093}{1993} \hepph{9209232}, \brk
S.~P.~Martin and M.~T.~Vaughn, \PLB{318}{331}{1993} \hepph{9308212}.


\bibitem{PDG:96}
Particle Data Group, \PRD{54}{1}{1996}.

\bibitem{KELLEY:92}
%
%
S.~Kelley, J.~L.~Lopez and D.~V.~Nanopoulos, \brk
\PLB{274}{387}{1992} \KEK{9111309}.


\bibitem{BARGER:93}
%
%
V.~Barger, M.~S.~Berger, P.~Ohmann, and R.~J.~N.~Phillips, \brk
\PLB{314}{351}{1993} \hepph{9304295}; \brk
%
%
\PRD{51}{2438}{1995} \hepph{9407273}.

\bibitem{BARDEEN:94}
%
%
W.~A.~Bardeen, M.~Carena, S.~Pokorski and C.~E.~M.~Wagner, \brk
\PLB{320}{110}{1994} \hepph{9309293}. \brk
M.~Carena, M.~Olechowski, S.~Pokorski, and C.~E.~M.~Wagner, \brk
\NPB{419}{213}{1994} \hepph{9311222}; \brk
\NPB{426}{269}{1994} \hepph{9402253}.

\bibitem{BANDO:94}
%
%
M.~Bando, K.-I.~Izawa and T.~Takahashi, \brk
\PTP{92}{143}{1994}  \hepph{9403284}; \brk
\PTP{92}{1137}{1994} \hepph{9408314}.


\bibitem{POKORSKI:95}
%
%
M.~Olechowski and S.~Pokorski, \brk
\PLB{344}{201}{1995} \hepph{9407404}, \brk
F.~M.~Borzumati, M.~Olechowski, and S.~Pokorski, \brk
\PLB{349}{311}{1995} \hepph{9412379}, \brk
H.~Murayama, M.~Olechowski and S.~Pokorski, \brk
\PLB{371}{57}{1996} \hepph{9510327}.


\bibitem{BANKS:88}
T.~Banks, \NPB{303}{172}{1988}.


\bibitem{HEMPFLING:94}
%
%
R.~Hempfling, \PRD{49}{6168}{1994}.



\bibitem{BANDO:96}
%
%
M.~Bando, Y.~Taniguchi and S.~Tanimura,  \brk
KU--AMP~96014, KUNS--1420, HE(TH)~96/15 \hepth{9610244}.

\bibitem{BARDEEN:90}
%
%
W.A.~Bardeen, C.T.~Hill and M.~Lindner, \PRD{41}{1647}{1990}.


%
%
%
%
%
%
%
%
%
%
%
%
%
%
%
%
%
%
%
%
%

\end{thebibliography}
\end{document}